\documentclass[prb,preprint,showpacs]{revtex4}
\usepackage{epsfig}
\usepackage{subfigure}
\begin{document}

\title{Structural and electronic properties of Si/Ge nanoparticles}

\author{Abu Md.\ Asaduzzaman\footnote{e-mail: a.asaduzzaman@mx.uni-saarland.de}
and Michael Springborg\footnote{e-mail: m.springborg@mx.uni-saarland.de}}
\affiliation{Physical and Theoretical Chemistry, University of  Saarland, D--66123 Saarbr\"ucken,
Germany}

\date{\today}

\begin{abstract}

Results of a theoretical study of the electronic properties of (Si)Ge and (Ge)Si
core/shell nanoparticles, homogeneous SiGe clusters, and Ge$\vert$Si clusters with an
interphase separating the Si and Ge atoms are presented. In general, (Si)Ge particles
are more stable than (Ge)Si ones, and SiGe systems are more stable than Ge$\vert$Si ones. It is
found that the frontier orbitals, that dictate the optical properties, are localized to
the surface, meaning that saturating dangling bonds on the surface with ligands may influence
the optical properties significantly. In the central parts we identify a weak tendency for
the Si atoms to accept electrons, whereas Ge atoms donate electrons. 

\end{abstract}


\maketitle

\section{Introduction}
\label{sec1}

Traditionally, materials properties have been controlled by varying structure and composition of the 
materials. During the last quarter of a century a new parameter has been added, i.e., size.
The fact that the materials properties change drastically when the dimension of the materials 
becomes comparable with the typical length scale of the phenomenon of interest together with 
the ability to control the production of materials in this size range has led to the development
of `nano-science'. 

One of the materials classes where materials in the nm range are expected to have a large 
impact on the development of new and/or better devices is semiconductors. For those, modified  
electronic properties may show up when the size of the nanocrystals is comparable with the 
spatial extension of the excitons. For instance, the light emission properties of semiconductor
nanocrystal quantum dots, specifically the tuning of color afforded by the quantum
size effect, is important for application of these materials in light emitting devices
\cite{ref01,ref02,ref03,ref04} and as biological fluorescence markers.\cite{ref05,ref06,ref07}

A further development which has opened up new possibilities to control and vary the 
materials properties is the successful production of 
core/shell nanoparticles, consisting of a core of one material coated by a well-defined
shell layer of another material. These systems exhibit unique and advanced properties over 
single-component nanoparticles, making them attractive for use in a wide range of real-world 
applications, and are therefore of extensive scientific and technological interest.
In the last few years, much effort has been focused on the synthesis, fabrication, and
characterization of the core/shell structured semiconductor heterostructures with 
tailored properties. The growth of the shell on the core
material to form a core/shell heterostructures has been successfully demonstrated on
the surface reconstruction of nanostructured material.
Among the ingredients that dictate the electronic, electrical,
optical, and chemical properties of core/shell nanostructures, the surface-to-volume 
ratio, the shell type and shell thickness are important, although a precise understanding
of the relations between structure, size, and composition on the one hand and property on
the other hand is lacking.

Recently, Lauhon {\sl et al.}\cite{ref08}  reported the epitaxial growth of crystalline
silicon-germanium and germanium-silicon core/sheath structures. Despite a 4\% lattice
mismatch of the macroscopic crystalline systems, the experimental study demonstrated that 
for Si core nanowire, the Ge shell is
fully crystallized at low temperatures and for a Ge core nanowire, an amorphous Si shell is
formed initially and after thermal annealing the shell becomes crystallized.
In another study, Malachias {\sl et al.} \cite{ref09} experimentally showed the 
formation of Ge domes
with a Si core and a Ge shell. Kolobov {\sl et al.}\cite{ref10} in their experimental study showed 
the formation of nanocrystals of a Ge core with a SiGe shell. On the other hand,
there are few experimental and theoretical studies on core/shell materials 
reported.\cite{ref11,ref12,ref13} Most studies on core/shell studies have 
considered binary compounds
like CdSe/ZnS, ZnS/CdS, CdSe/CdS, etc. Core-shell studies on pure elements are very scarce.
Recently, Musin and Wang\cite{ref14} reported a theoretical study
on the epitaxial Si-Ge core/shell structure, a theoretical realization of the experimental work of
Lauhon {\sl et al.}

\begin{figure}
\begin{center}
\subfigure{\includegraphics[scale=0.40]{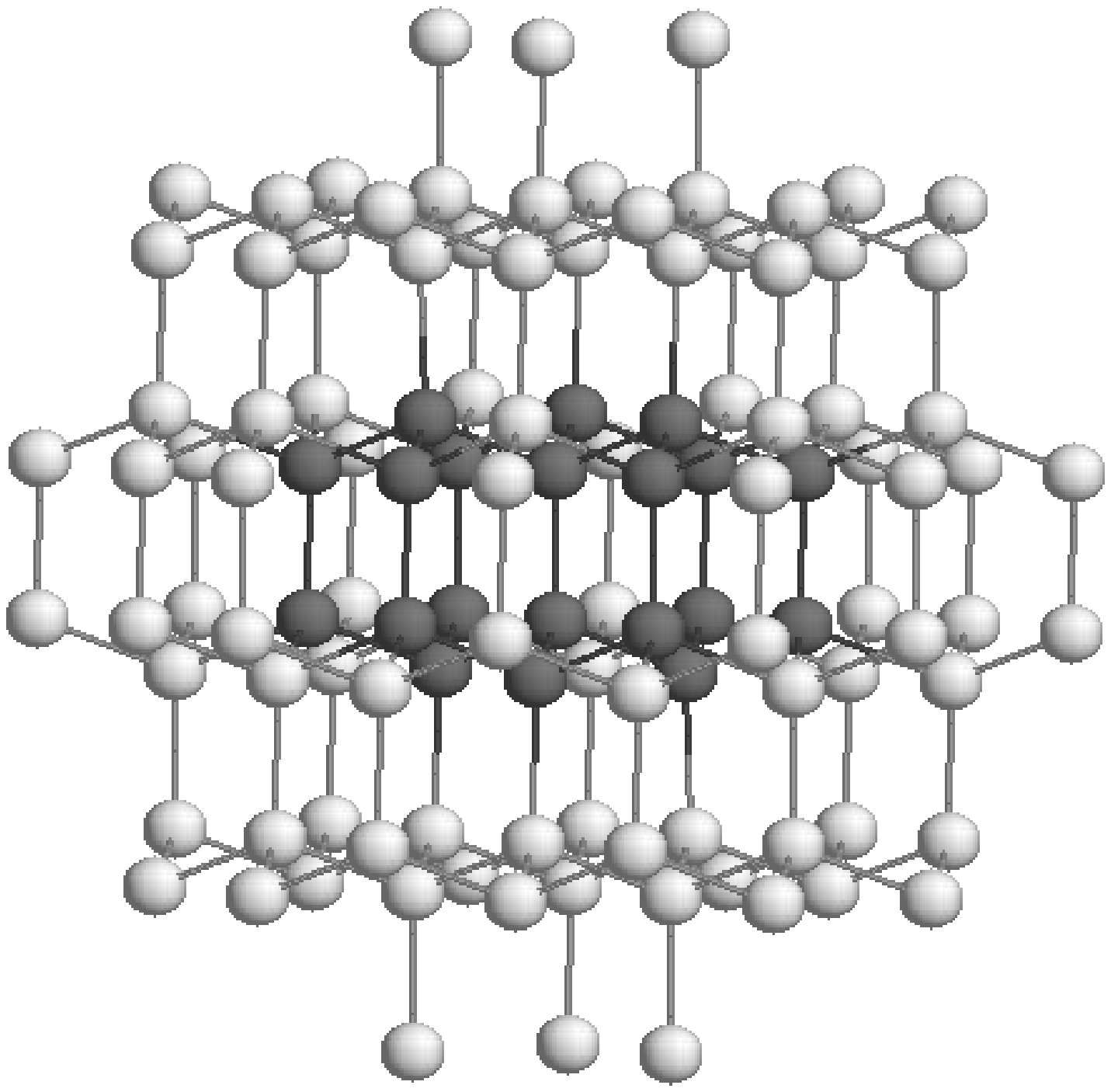}}
\subfigure{\includegraphics[scale=0.40]{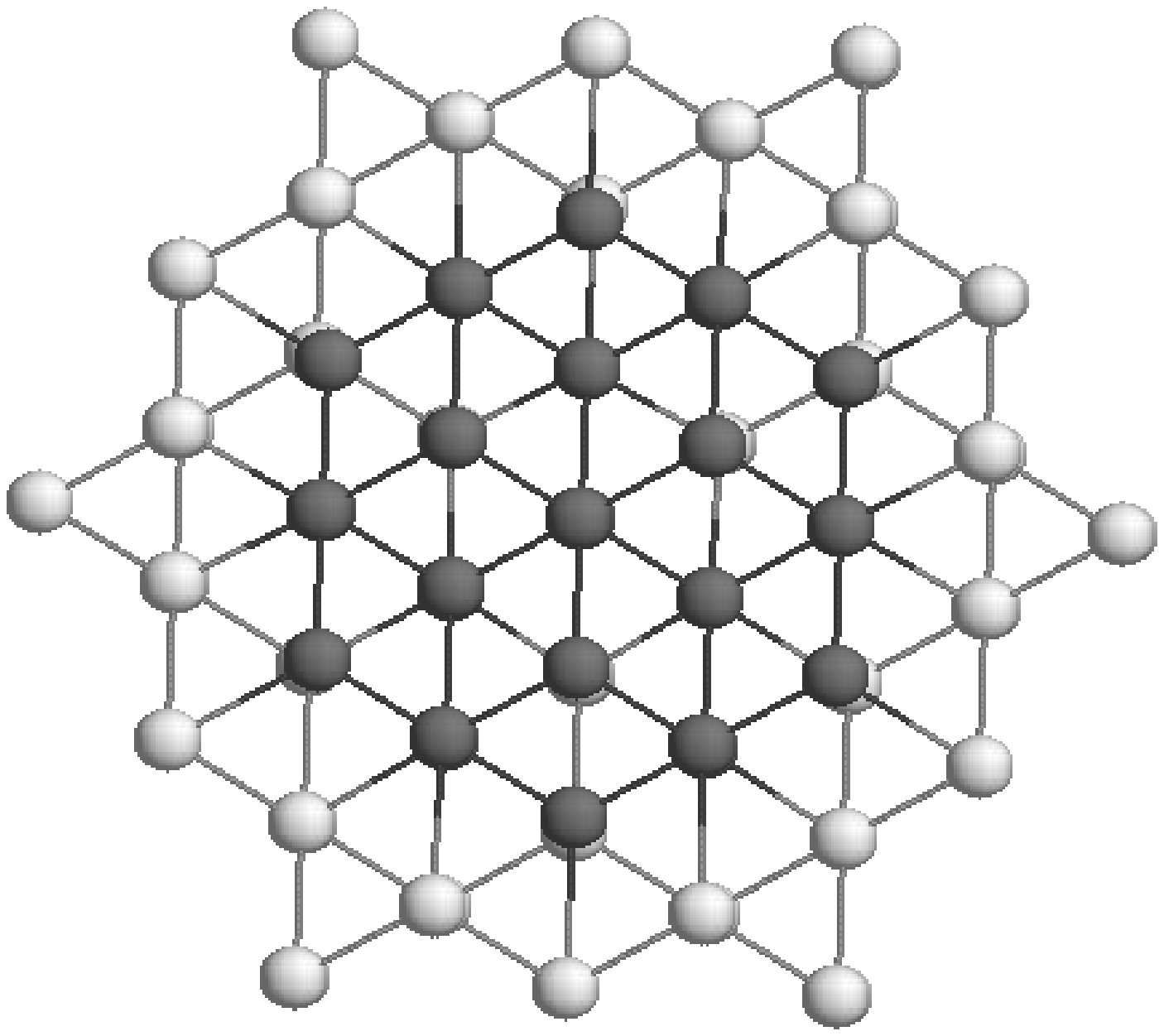}}
\subfigure{\includegraphics[scale=0.40]{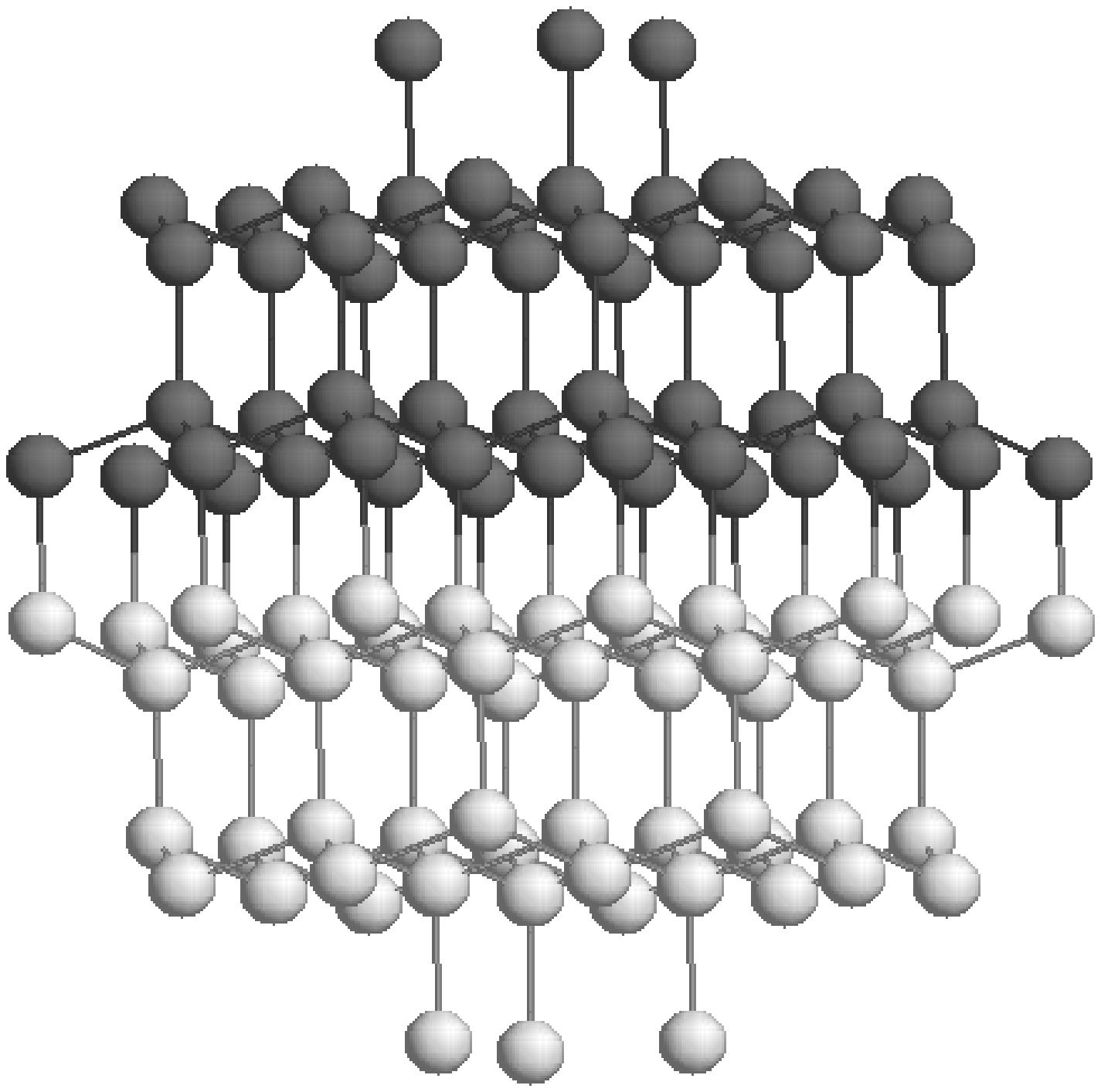}}
\subfigure{\includegraphics[scale=0.40]{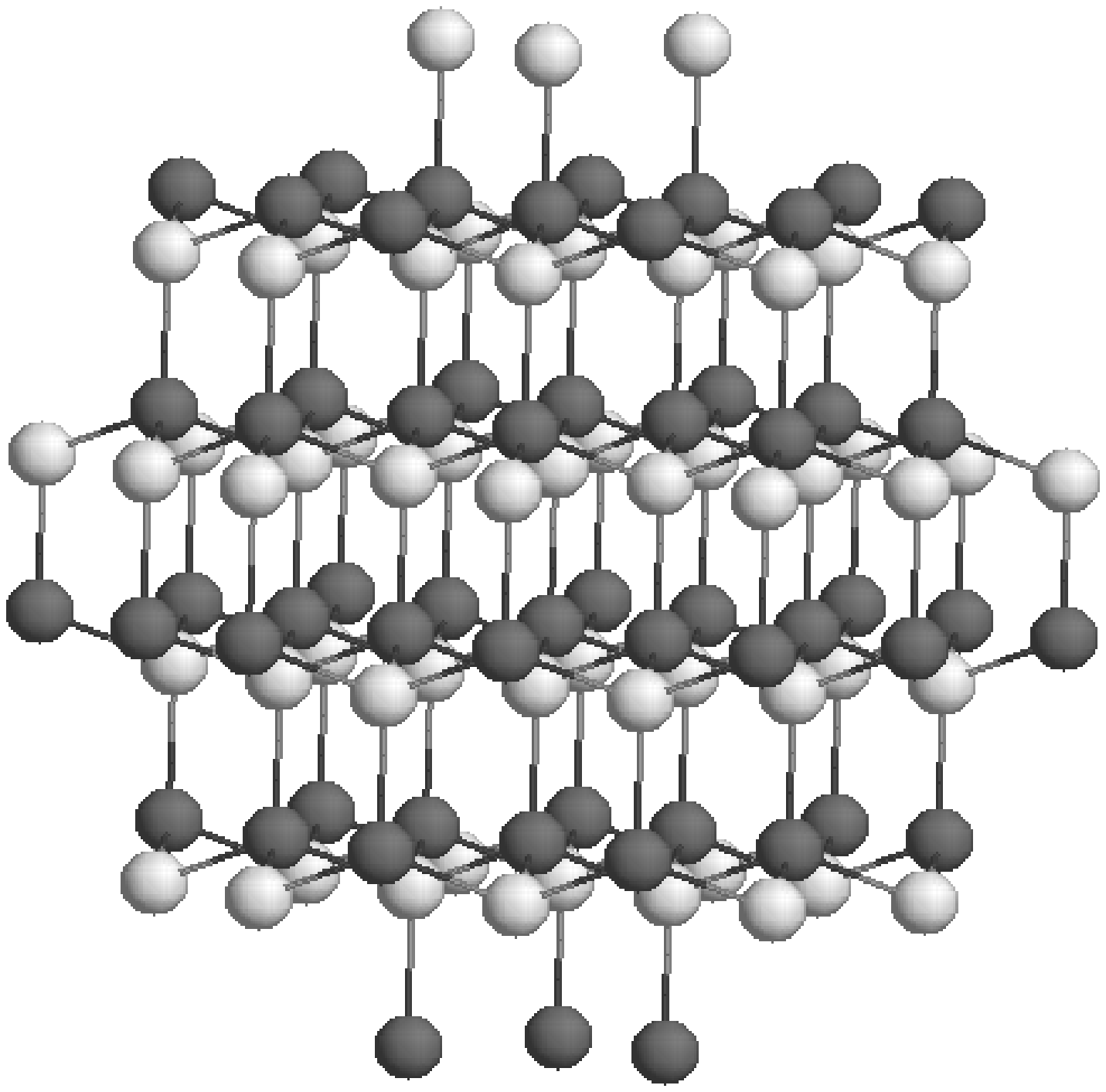}}
\end{center}
\caption{Ball and stick representation of a representative core/shell structure (upper panel). More
and less dark parts
represent the core and the shell, respectively. Both the sideview (left panel) and 
the cross-sectional view (right panel) are presented. Si$\vert$Ge (left panel) and
homogeneous SiGe (right panel) systems are represented in the lower panel.}
\label{fig01}
\end{figure}

In this paper, we present the results of a theoretical study of 
structural and electronic properties of naked Si-Ge and Ge-Si core/shell nanoparticles.
Musin and Wang  took into consideration the epitaxial growth
of Si-Ge core/sheath nanowires and studied the compositional dependency of the structural 
parameters and the band gap energy. Here, we have considered naked 
Si-Ge and Ge-Si core/shell nanoparticles for which we assumed that the structure is 
related to that of a spherical cut-out of the infinite crystal with a zincblende- or diamond-like 
structure. Although our assumption may affect the results of the calculations, we believe that by 
considering a larger number of sizes (almost 100 sizes with up to in total almost 200 atoms) our
study allows for drawing general conclusions on those systems. Starting with a small core
of only 8 atoms and a thin shell of 24 atoms we have gradually increased the sizes of both
core and shell. 
In order to find out how the properties of core/shell particles depend on the two elements,
we have, in addition, carried out calculations on pure Si clusters, pure Ge clusters, 
homogeneous SiGe clusters, and spherical structures with half of the sphere made up of Si and
the other half made up of Ge atoms (Si$\vert$Ge).  
A representative example of a core/shell particle as well as of homogeneous SiGe and 
of a Si$\vert$Ge particle is
depicted in Fig.\ \ref{fig01}. We mention that the experimentally studied systems are considerably
larger (containing up to several 1000s of atoms), which is very
difficult to study theoretically. Therefore, in this study we have limited ourselves
to a detailed study of different core/shell structures along with pure Si and Ge, homogeneous SiGe, and
Si$\vert$Ge systems with up to around 200 atoms, which
is around 2 nm in radii. We have optimized all the structures to their nearest local total-energy minima
whereby all atoms were allowed to move.

\section{Computational outline}
\label{sec2}

We have used a parameterized density-functional tight-binding method that has been
described in detail elsewhere.\cite{ref15,ref16,ref17} The total energy relative to the 
isolated atoms of a given system is written by
\begin{equation}
E_{b}=\sum_{i}\epsilon_{i}-\sum_{jk}\epsilon_{jk}+\frac{1}{2}\sum_{k\neq l}U_{kl}(\vert\vec R_{k}
-\vec R_{l}\vert)
\end{equation}
Here, $\epsilon_{i}$ is the energy of the $i$th orbital for the system of interest and
$\epsilon_{jk}$ is the energy of the $j$th orbital for the isolated $k$th atom. $U_{kl}$ is
a pair potential between the $k$th and $l$th atom. The valence single-particle eigenfunctions
$\psi_{i}(\vec r)$ to the Kohn-Sham equation are expanded in a set of atom-centered basis
functions $\chi_{klm}(\vec r)$, where $k$ denotes the atom and $(l,m)$ the angular
dependence. The effective one-electron potential in the Kohn-Sham Hamiltonian is
approximated as a superposition of the atomic potentials of the corresponding neutral atoms, and
it is assumed that the matrix elements $\langle\chi_{k_1l_1m_1}\vert V_j\vert \chi_{k_2l_2m_2}\rangle$
(with $V_j$ being the atomic potential at atom $j$) are vanishing unless $k_1=j$ or $k_2=j$. 
Thus, only two-center Hamiltonian matrix elements are considered and calculated exactly
within the Kohn-Sham basis for the diatomic molecules. Finally, the pair potentials $U_{kl}$
are determined so that the binding energy curve of the diatomics are well reproduced.
Only the $3s$ and $3p$ functions of Si and the $4s$ and $4p$ functions of Ge were 
explicitly included in the calculations, 
whereas all other electrons were treated within a frozen-core approximation.

It is obvious that, the approach we are using has been designed for the smallest possible
systems Si$_2$, Ge$_2$, and SiGe. We have verified the capability of this method to do 
calculations
for larger systems by studying some characteristics of bulk Si and Ge. 
The experimental lattice constants of crystalline Si and Ge are 5.43 and 5.66 \AA, respectively,
whereas our calculations give 5.46 and 5.71 \AA, respectively, i.e.,
within less than 1\% of the experimental values. 
The experimental band gaps of bulk Si and Ge are 1.12 and 0.66 eV, respectively, and our
calculated values are 1.097 and 0.65 eV, respectively which are also close to the
experimental values. Here, the standard problem of density-functional calculations to yield
too small band gaps seems to be absent, mainly due to the fact that our basis set is minimal in size.

In addition to the clusters containing both Si and Ge atoms, we have also studied small clusters
of pure Si or Ge, separately. Among the different arrangements of three atomic Si and Ge clusters, 
clusters with $C_{2v}$ symmetry are those 
of the minimum energy configuration. Clusters with $D_{2h}$ symmetry are the minimum energy structures 
among the four atomic clusters 
for both Si and Ge. Earlier theoretical studies on Si and Ge\cite{ref18} and on Si\cite{ref19}
also reported those minimum energy structures. 

In all cases we relaxed initial structures that we constructed by cutting out a spherical part
of a diamond-like crystal. The initial lattice constant was taken as that of the pure crystalline
systems (for the pure Si and Ge clusters) or as the average of those two values (for the 
heteroatomic structures). The center of the sphere was taken as the midpoint of a nearest-neighbor
bond, giving that the number of atoms would be 2, 8, 20, 32, 38, 56, 74, 86, 116, 130, 166, or
190 if 1, 2, $\dots$, 12 atomic shells were included in the initial structure (here, we define an
atomic shell as being the set of atoms that has the same distance to the center of the 
spherical cut-out in the initial structure). Subsequently, the
initial structure was allowed to relax to its closest total-energy minimum, whereby all atoms 
(i.e., both in the inner part and in the surface region) were displaced until the forces on them
vanish.

\section{Results}
\label{sec3}

We studied in total 95 different structures. Each of those structures consists of $N_{{\rm Si},i}$
Si atoms and $N_{{\rm Ge},i}$ Ge atoms, $i=1,2,\dots,95$. Using a least-squares fit we approximated 
the binding energy of those 95 structures by a sum of atomic energies,
\begin{equation}
E_{b,i}\simeq E_{\rm Si}N_{{\rm Si},i}+E_{\rm Ge}N_{{\rm Ge},i}\equiv\tilde E_{b,i}.
\label{eqn01}
\end{equation}
Subsequently, we defined one stability energy for each cluster,
\begin{equation}
\Delta E_1 =E_{b,i}-\tilde E_{b,i}
\end{equation}
which is the more negative the more stable the cluster is. Finally, we analyse this quantity per atom,
i.e., 
\begin{equation}
\Delta E_1 / N = \Delta E_1 / (N_{{\rm Si},i}+N_{{\rm Ge},i}).
\end{equation}
We also considered the stability energy
\begin{equation}
\Delta E_2 / N = E_{b,i} / (N_{{\rm Si},i}+N_{{\rm Ge},i}).
\end{equation}

Each of those quantities is analysed as a function of the number atomic shells either in the core, in the 
shell, or in the complete nanostructure. I.e., with $N_t$ being
the total number of atomic shells, $1\le N_t\le 12$, for the core/shell particles we have $N_c$
atomic shells in the core and $N_s=N_t-N_c$ shells in the shell part, whereas $N_s=N_t, N_c=0$ for the 
homogeneous SiGe clusters and for the Si$\vert$Ge nanoparticles. 

The fit of Eq.\ (\ref{eqn01}) resulted in $E_{\rm Si}=$ -2.37 eV and $E_{\rm Ge}=$ -3.58 eV. The 
fact that $E_{\rm Ge}$ is more negative than $E_{\rm Si}$ implies that it is energetically more
favorable for Ge atoms than for Si atoms to be incorporated into those nanostructures, although
the difference in the two energies is relatively small. On the other hand, the cohesive energy
of the solids equals 4.63 and 3.85 eV/atom for Si and Ge, respectively (see, 
e.g., [\onlinecite{ref18a}]), giving that for the elemental solids it is energetically more 
favorable for Si than for Ge atoms to be incorporated into the solids. The difference between
our results and those for the elemental solids may be due to the differences in the systems (both
with respect to composition and regarding size) and in the finite number of structures that we 
include in our fit. 

The fact that $E_{\rm Ge}<E_{\rm Si}$ also means that when comparing 
$\Delta E_1$ and $\Delta E_2$, the larger the number of Ge atoms is compared with that of Si atoms,
the lower is $\Delta E_2$ compared with $\Delta E_1$. 

\begin{figure}[tbp]
\begin{picture}(15,20)
\put(2,0){\psfig{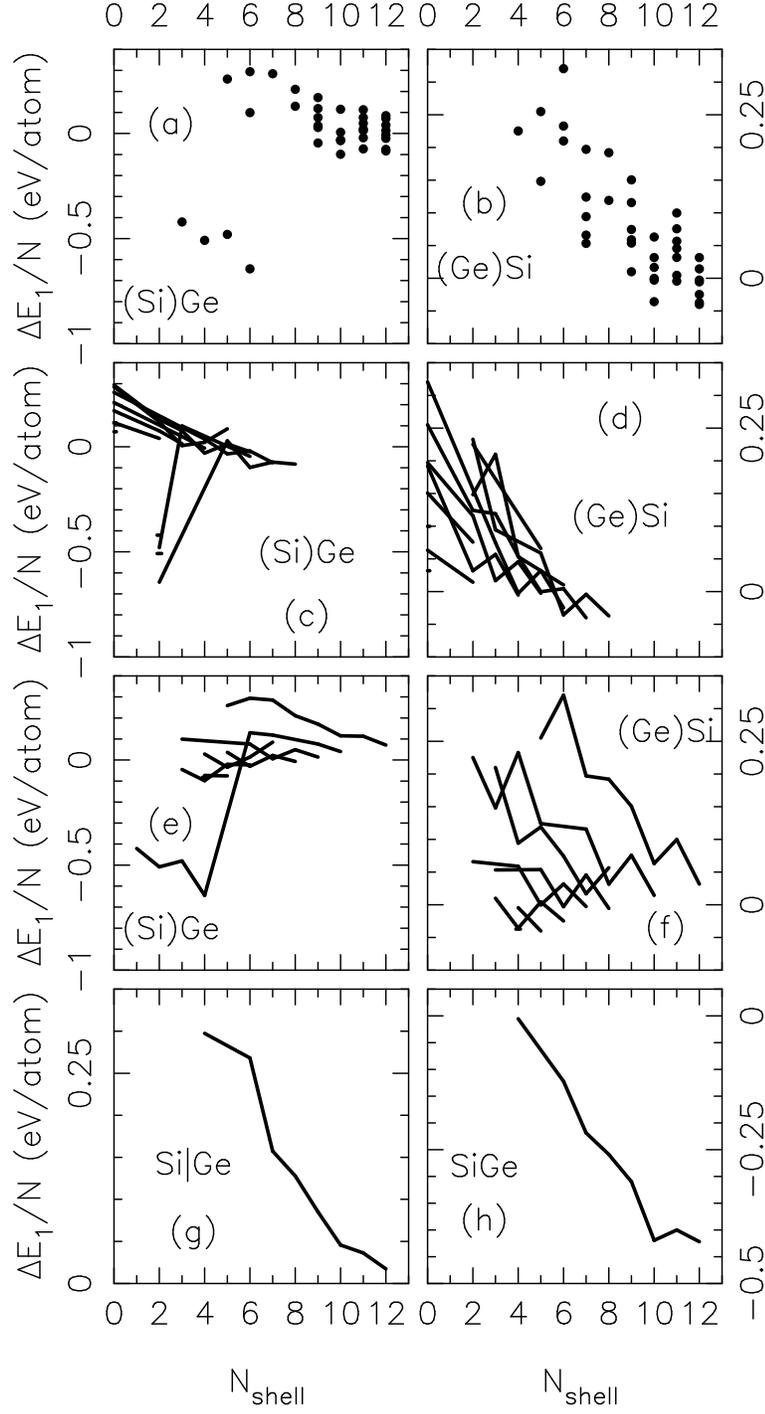}}
\end{picture}
\caption{The variation in the stability energy per atom $\Delta E_1/N$ as a function of the number 
of atomic shells for the different systems, as indicated in the panel. $N_{\rm shell}$ is the
total number of shells, $N_t$, in (a), (b), (g), and (h), the number of shells in the core, $N_c$,
in (e) and (f), and the number of shells in the shell $N_s$, in (c) and (d). Finally, the lines 
in (c)--(f) connect the values for the systems with the same number of atomic shells in (c,d) the
core or in (e,f) the shell.}
\label{fig01a}
\end{figure}

\begin{figure}[tbp]
\begin{picture}(15,20)
\put(2,0){\psfig{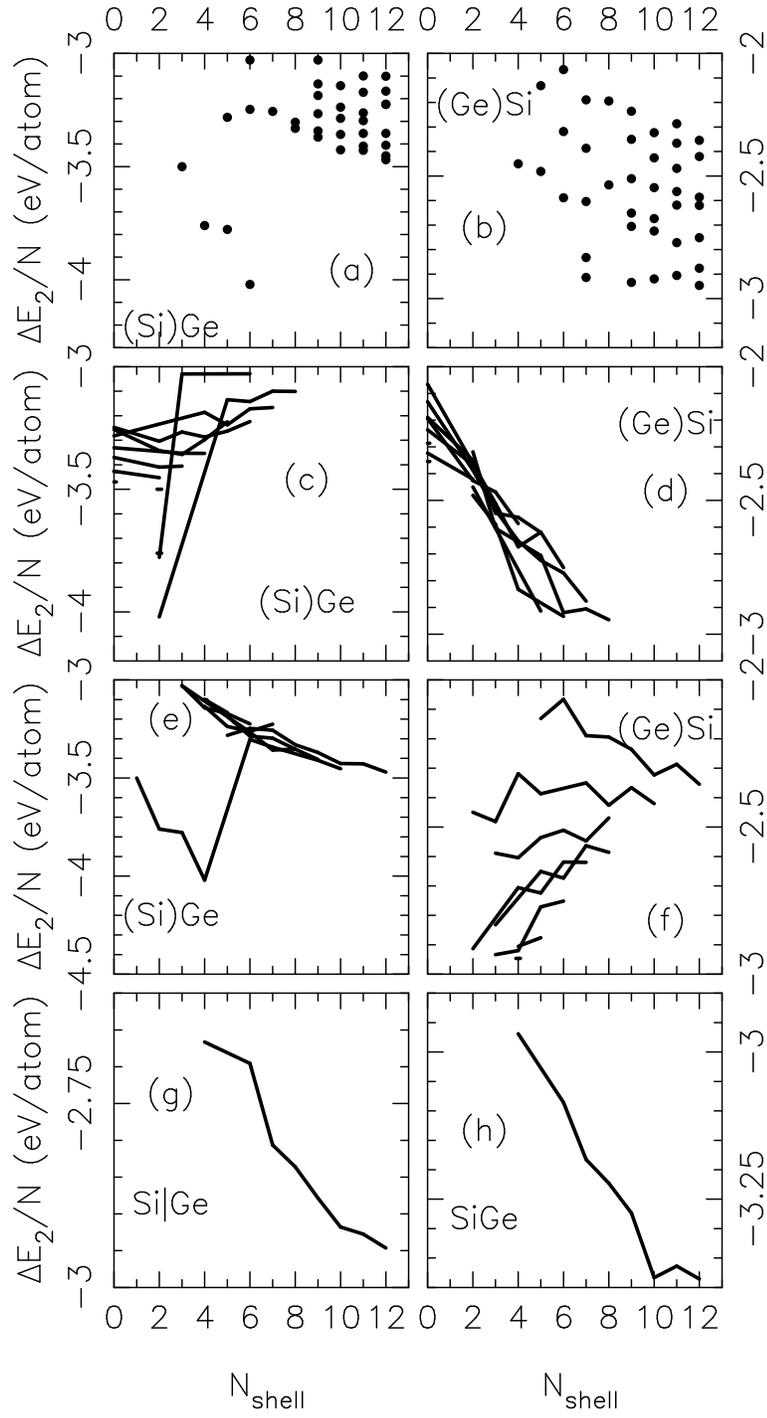}}
\end{picture}
\caption{As Fig.\ \ref{fig01a}, but for the stability energy per atom $\Delta E_2/N$.}
\label{fig01b}
\end{figure}

In Figs.\ \ref{fig01a} and \ref{fig01b} we show the two energies $\Delta E_1/N$ and $\Delta E_2/N$
as a function of the number of shells in either the core region, in the shell region, or in the
complete cluster. 

First, the results for the Si$\vert$Ge and the homogeneous SiGe nanosystems show clearly that 
the latter is more stable than the former. In addition, in both cases the binding energy per atom is
a decaying function of the size of the system. This behavior is often found for clusters, and 
can be related to the presence of the surface atoms for which the appearance of dangling bonds and
lower coordinations lead to energetically less favorable situations. When the system
size is increased, the relative importance of the surface is reduced, leading to an overall 
decaying total energy per atom as a function of size. The fact that the precise number of
surface atoms is not completely regular as a function of cluster size can 
explain the deviations from a completely smooth behavior.

Comparing Figs.\ \ref{fig01a}(b) and \ref{fig01b}(b) the finding that the latter is fairly unstructured,
whereas the former shows a clear decaing behavior as function of size suggests that for the (Ge)Si systems,
i.e., a Ge core covered with Si shells, the total energy for this system depends only marginally
on its size. 

On the other hand, Figs.\ \ref{fig01b}(d) shows that for the (Ge)Si nanoparticles the variation
in the total energy is mainly determined by the number of atomic Si shells in the shell part and
largely independent of the number of atomic Ge shells in the core region. A similar result is
found for the (Si)Ge nanoparticles in Fig.\ \ref{fig01a}(c) with, however, some few
exceptions that are recovered in all results for these systems, i.e., in  
Figs.\ \ref{fig01a}(a,c,e) and Figs.\ \ref{fig01b}(a,c,e). It turns out that these exceptions 
are among the 
smallest systems we have studied with only two atomic shells of Si covered with a low number
of Ge shells. We mention that the systems with no atomic Si shells of Si, are less stable, 
so that the (Si)Ge nanoparticles with two atomic shells of Si covered
with a small number of Ge shells indeed are the most stable systems of the present study. We do not
have a precise explanation for this finding. 

A further result can be seen when comparing the (Si)Ge core/shell systems with the (Ge)Si ones:
the former are more stable than the latter. Since the (Si)Ge systems contain a Ge surface, whereas
the (Ge)Si systems contain a surface of Si atoms, this result may be a simple consequence of the lower 
surface energy of Ge compared with that of Si (see, e.g., [\onlinecite{jap06}]). 

\begin{figure}[tbp]
\begin{picture}(15,20)
\put(2,0){\psfig{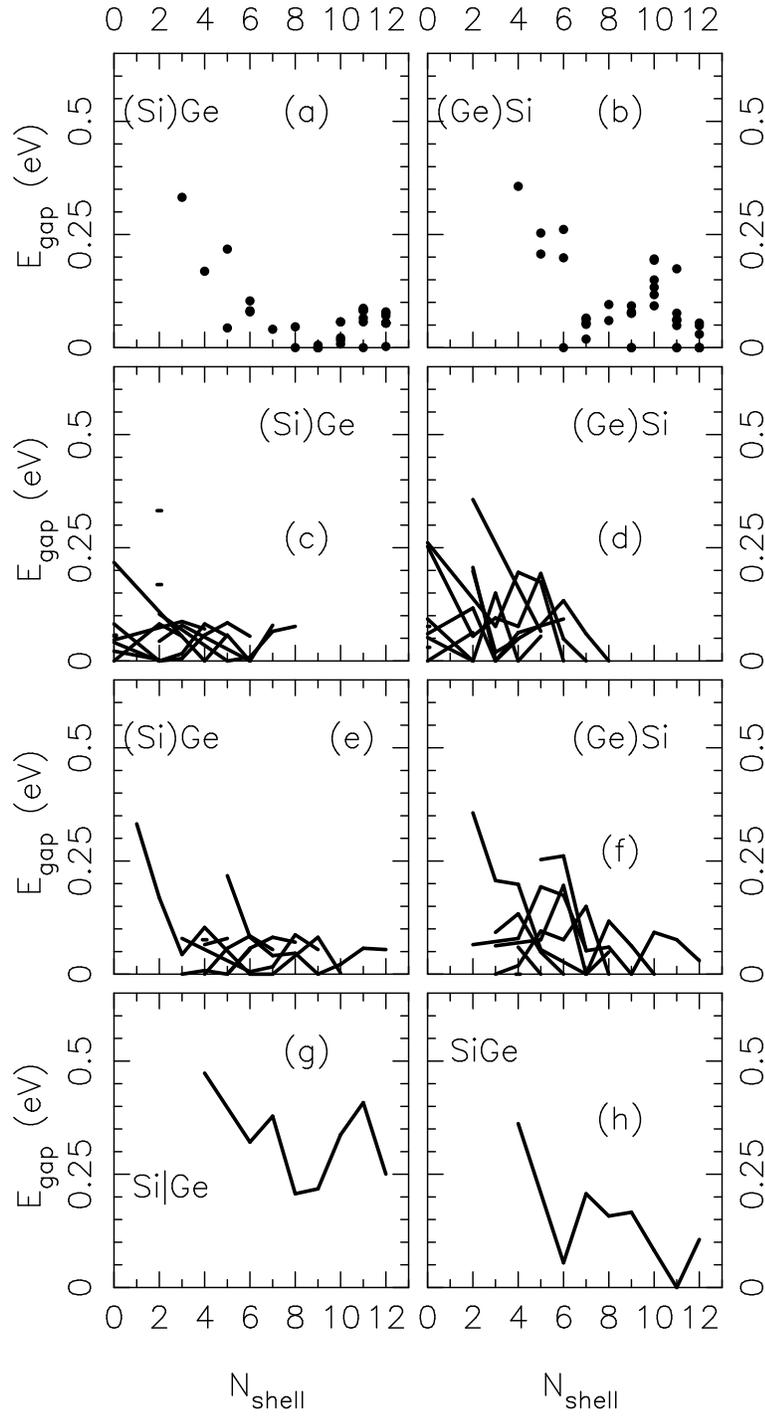}}
\end{picture}
\caption{As Fig.\ \ref{fig01a}, but for the HOMO-LUMO energy gap, $E_{\rm gap}$.}
\label{fig01c}
\end{figure}

The HOMO-LUMO gap, $E_{\rm gap}$, 
(i.e., the energy gap between the highest occupied molecular orbital and the
lowest unoccupied molecular orbital) is shown in Fig.\ \ref{fig01c} similarly to Figs.\ \ref{fig01a} 
and \ref{fig01b}.
With no exception the gap is significantly smaller than those of the elemental solids and
in many cases it is even close to vanishing. The systems with the largest gaps are the 
core/shell nanoparticles with the smallest number of atomic shells in the complete system as well
as the Si$\vert$Ge systems. 
Marsen {\sl et al.},\cite{ref22,ref23} in their experimental
study, showed that the band gap of small silicon clusters is larger (the largest value being 
0.45 eV) than that of the large clusters (most clusters have gap ranging from 0.025 to 0.10 eV) 
and the gap is size dependent. Besides that, they observed essentially no correlation between size of the 
gap and size of the cluster, which is in good agreement with our results. 

\begin{figure}[tbp]
\begin{picture}(15,10)
\put(2,0){\psfig{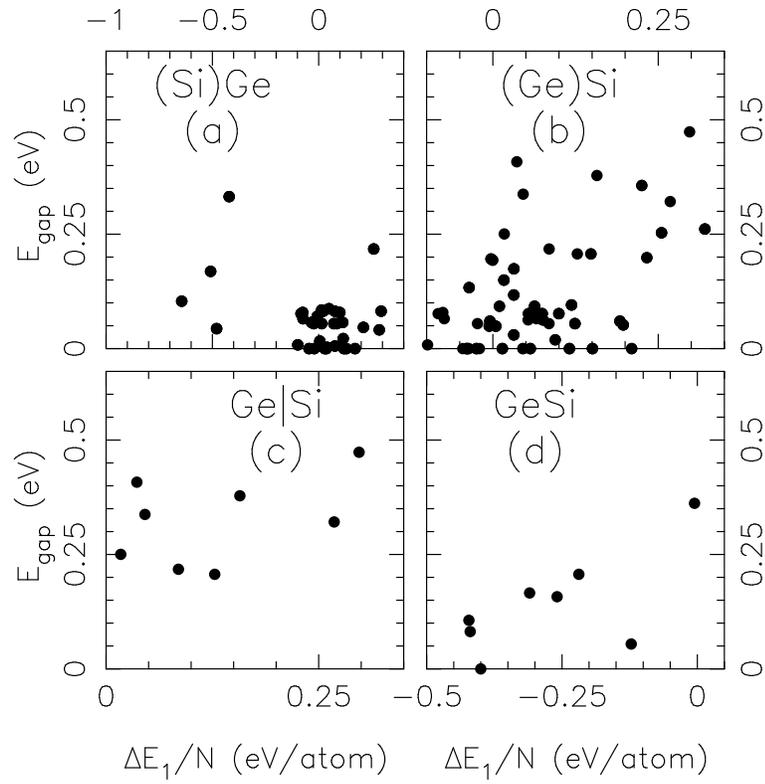}}
\end{picture}
\caption{The HOMO-LUMO energy gap as a function of the stability energy per atom
$\Delta E_1/N$ for (a,b) the core/shell systems with (a) Ge covering Si and (b) Si
covering Ge as well as for (d) the homogeneous GeSi systems and (c) the Ge$\vert$Si 
systems.}
\label{fig01d}
\end{figure}

In an earlier work on a number of stoichiometric II-VI and III-V nanoparticles we found a close
correlation between the HOMO-LUMO gap and the stability.\cite{ref18b} In order to explore whether
a similar correlation exists here, we show in Fig.\ \ref{fig01d} $E_{\rm gap}$ as a function of
$\Delta E_1/N$. It is clear that no correlation is found in the present case. Replacing $\Delta E_1/N$
by $\Delta E_2/N$ does not change this conclusion.

\begin{figure}
\centering
\epsfig{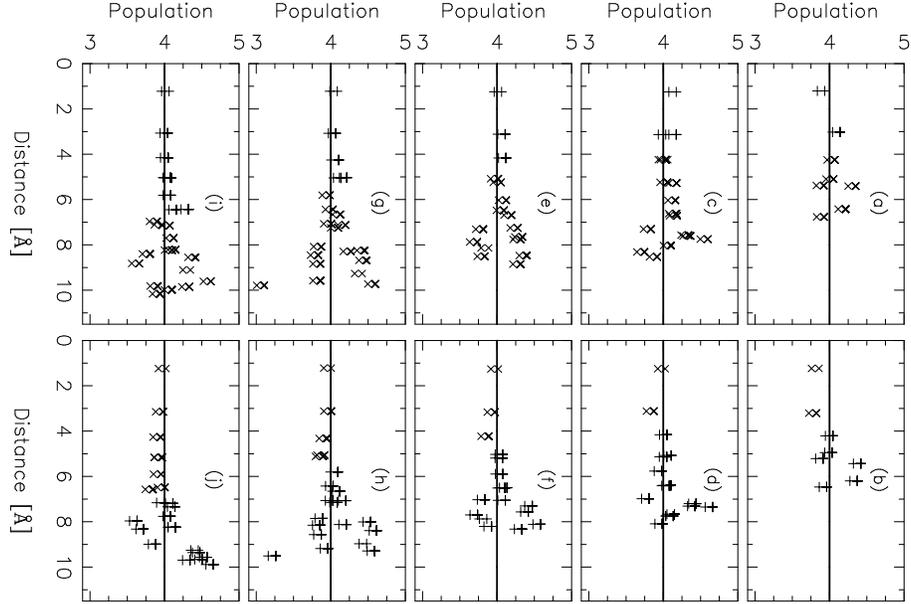}
\caption{Radial distribution of Mulliken gross populations of valence electrons
of Si ($+$) and Ge ($\times$) for (left part) (Si)Ge and (right part) (Ge)Si core/shell
particles with $n_c$ atoms in the core and $n_s$ atoms in the shell. $(n_c,n_s)$ equals
(a),(b) (8,48), (c),(d) (8,108), (e),(f) (20,110), (g),(h) (32,134), and (i),(j) (56,134). The
horizontal solid line marks the value (4) for neutral Si and Ge atoms.}
\label{fig03}
\end{figure}

Ge/Si nanowires with a Si sheath covering a Ge core have been of some interest recently (see, e.g.,
[\onlinecite{nat06}]), also as active components in semiconductor devices. In these, it is 
assumed that a hole gas is formed in the Ge wire. If the systems were isolated, this would imply
a net electron transfer from Ge to Si, which, then, also may occur for the system of the present
study. In order to study this, we first determine the center of the cluster with $n$ Si atoms
and $m$ Ge atoms,
\begin{equation}
\vec R_{0}=\frac{1}{n+m}\sum^{n+m}_{j=1}\vec R_{j},
\end{equation}
and, subsequently, for each atom its so-called radial distance
\begin{equation}
r_{j}=\vert\vec R_{j}-\vec R_{0}\vert.
\end{equation}
Subsequently, we plot the Mulliken gross populations of the individual atoms as a function of
the radial distance; cf.\ Fig.\ \ref{fig03}. Indeed, it is seen that for the (Si)Ge core/shell
particles there is a slight tendency for the atomic populations on the Si atoms to be larger than
4, whereas for the (Ge)Si core/shell particles, the Ge atomic populations are on the average
below 4. On the other hand, in the shell region, the trend is less clear. However, in total, 
these core/shell systems do show tendency towards electron- or hole-gas formation in the 
central part, depending on whether Si or Ge is the material of the core. 
The finding that the atomic populations deviate most from the value of 4
for the atoms closest to the surface is also observed for the Si$\vert$Ge systems. 
For the homogeneous SiGe systems we also observe a small electron transfer from Ge to Si
in the central part, which, however, not is the case for the central parts of the Ge$\vert$Si
systems.

Much of the interest in semiconductor nanoparticles is connected with their partly controllable
optical properties that first of all are dictated by the properties of excitons, which
in turn are determined by the energy and the charge distribution of the orbitals
closest to the Fermi level, i.e., the HOMO and the LUMO. Optical relaxation processes
strongly depend on the spatial distribution of these orbitals. Therefore, we shall study these.
To this end we construct for any orbital the density
\begin{equation}
\rho_{i}(\vec r)=\sum_{j}N_{ij}\left(\frac{2\alpha}{\pi}\right)^{3/2}\exp[-\alpha(\vec r-
\vec R_{j})^2].
\end{equation}
Here, $N_{ij}$ is the Mulliken gross population for the $j$th atom and $i$th orbital, and
$\alpha$ is chosen `reasonably', so that illustrative figures result.

\begin{figure}
\centering
\epsfig{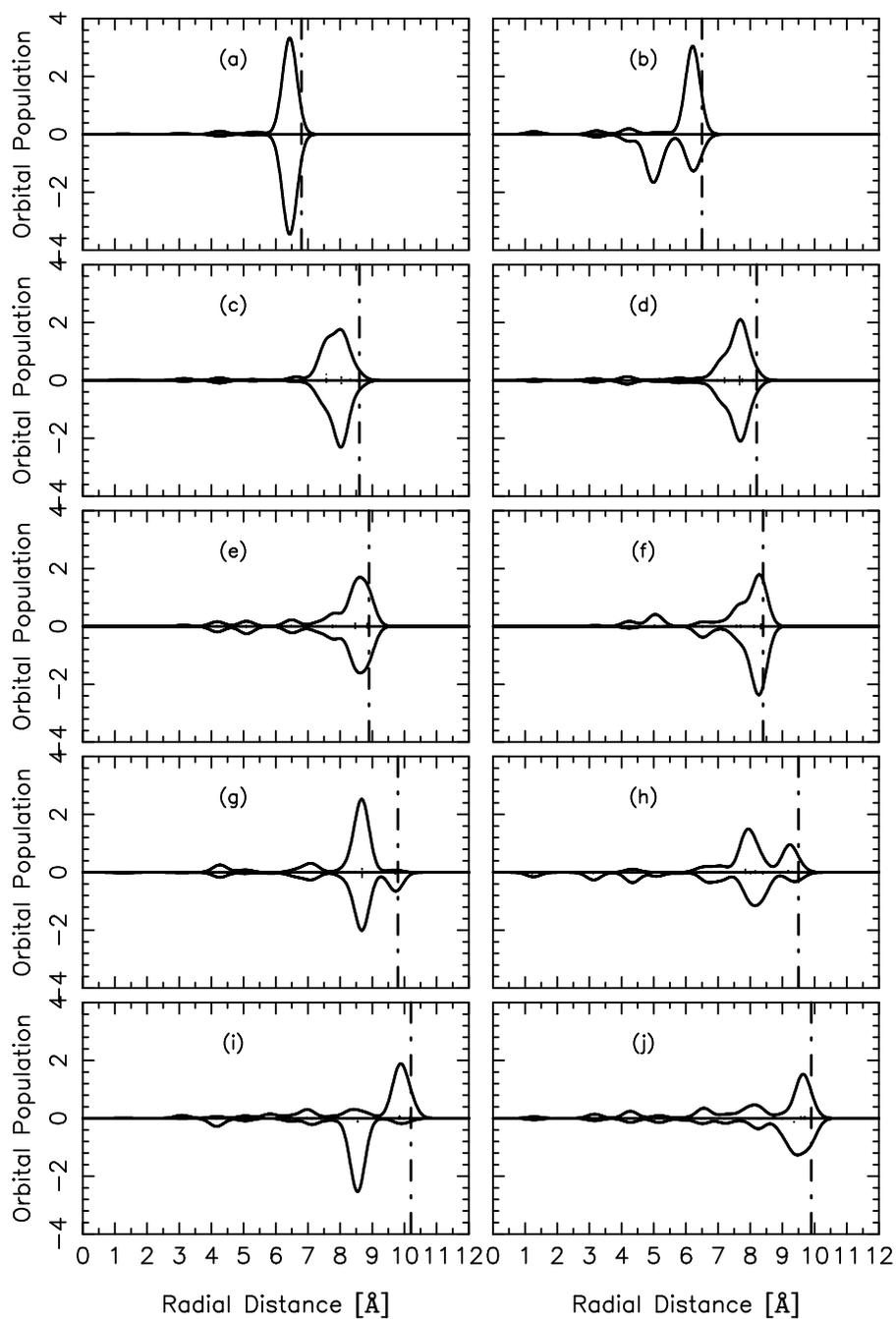}
\caption{Schematic presentation of the radial electron distribution of the HOMO (curves
pointing upward) and the LUMO (curves pointing downward) for the same systems as in
Fig.\ \ref{fig03}.}
\label{fig04}
\end{figure}

Fig.\ \ref{fig04} shows the resulting 
schematic representation of this radial dependence of the HOMO and LUMO. It is seen 
that in most cases, both the HOMO and the LUMO are localized to the surface
of the clusters, irrespectively of whether we consider (Si)Ge or (Ge)Si systems. In an earlier
study on CdSe/CdS core/shell nanoparticles \cite{ref21} it was found that one may find
systems for which the LUMO and the HOMO were localized in different parts (i.e., in the 
shell and in the core) of the system, but this is obviously not the case for the present 
systems. Since the gap of crystalline Ge is smaller than that of crystalline Si, and since
the HOMO and LUMO of the present systems are located to the surface region, it may be suggested
that the gap of the (Si)Ge particles is smaller than that of the (Ge)Si systems. Actually, such
a difference has been found for the pure Ge and Si systems by Melnikov and Chelikowsky.\cite{ref24}
However, in the present study such a trend is only very marginally found for the core/shell
systems, which is in agreement with the theoretical results of Musin and Wang\cite{ref14} who
studied Ge/Si and Si/Ge core/sheath nanowires. Moreover, Musin and Wang found a larger band gap 
than those of the pure, crystalline elements for all systems, which most likely is due to the 
passivation of the surface included in their study. This once again indicates that without 
passivation the frontier orbitals are localized to the surface region.

\begin{figure}
\centering
\epsfig{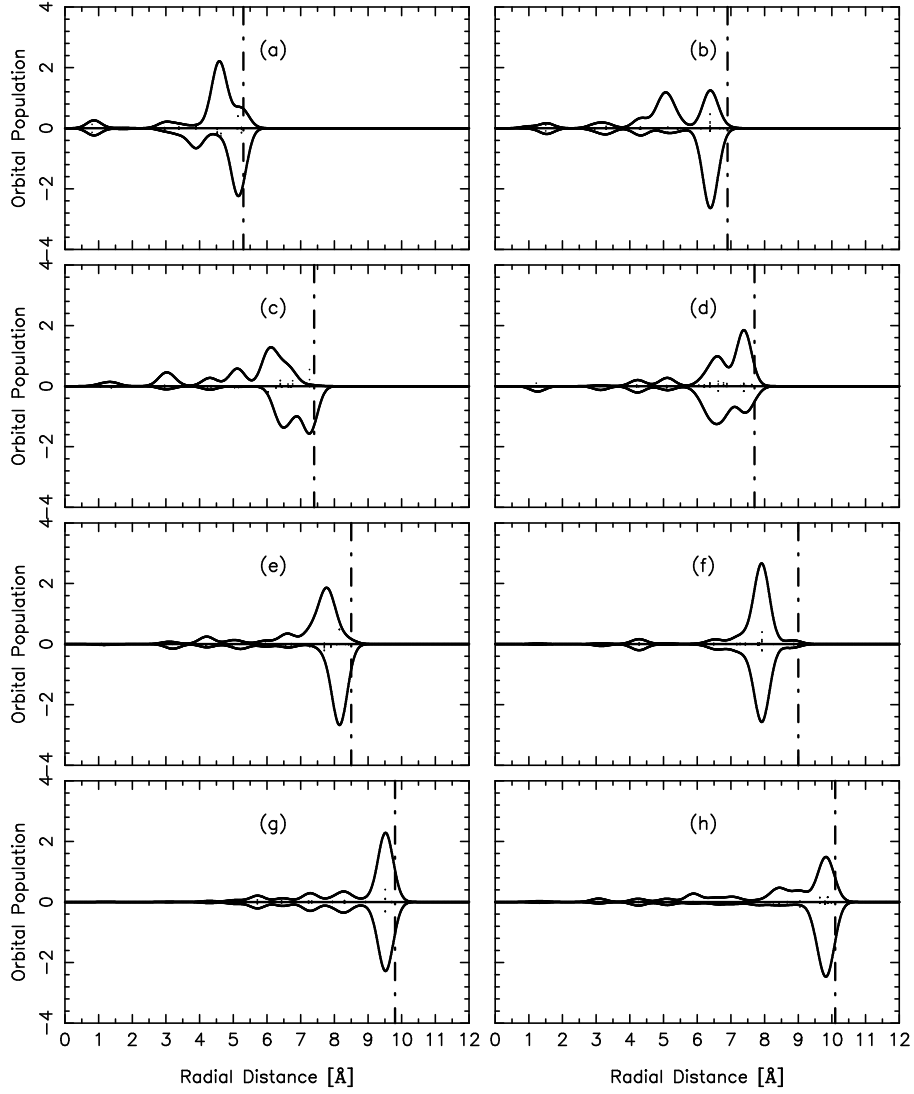}
\caption{Schematic presentation of the radial electron distribution of the HOMO (curves
pointing upward) and the LUMO (curves pointing downward) for homogeneous SiGe nanoparticles 
with (a) 32, (b) 56, (c) 74, (d) 86, (e) 116, (f) 130, (g) 166, and (h) 190 atoms.}
\label{fig05}
\end{figure}

\begin{figure}
\centering
\epsfig{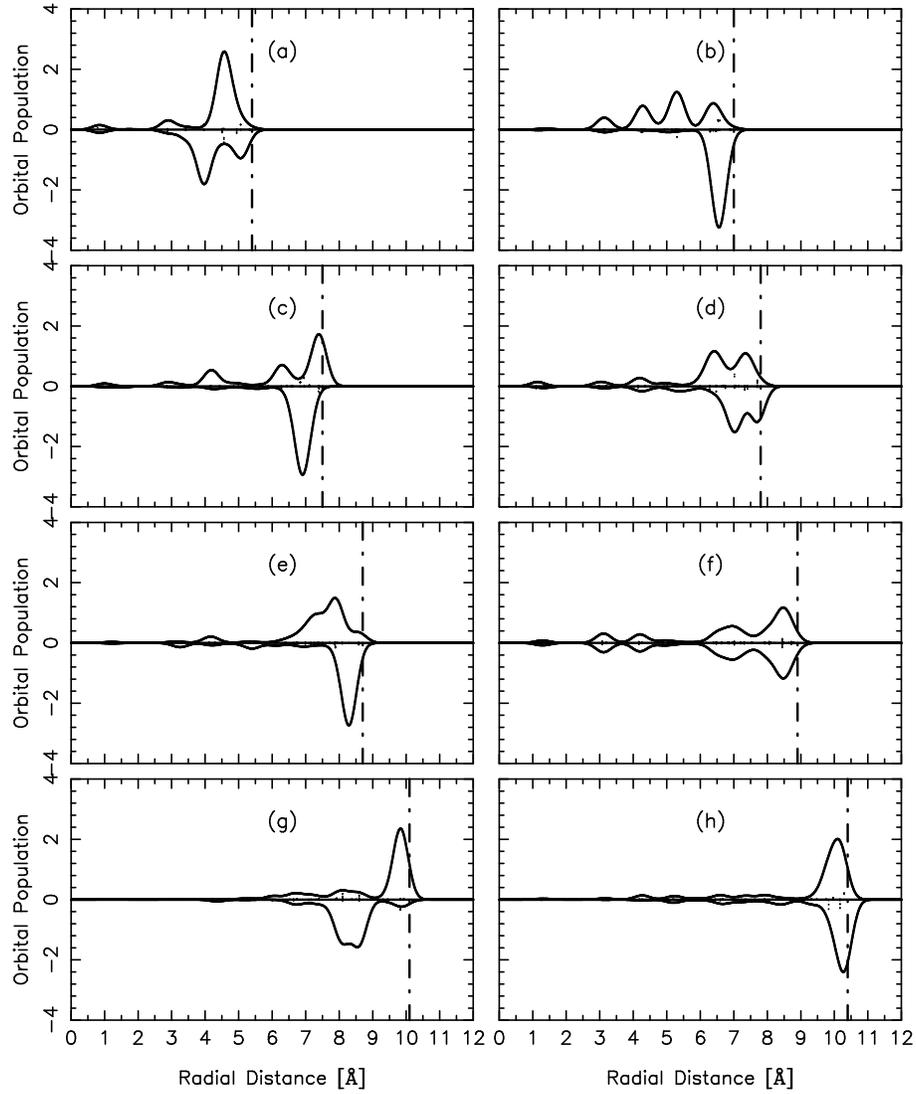}
\caption{Schematic presentation of the radial electron distribution of the HOMO (curves
pointing upward) and the LUMO (curves pointing downward) for the same number of atoms as
Fig.\ \ref{fig05} but for Si$\vert$Ge nanoparticles.} 
\label{fig06}
\end{figure}

Also for the homogeneous SiGe systems the frontier orbitals are located to the surface, as
can be seen in Fig.\ \ref{fig05}. For the Si$\vert$Ge systems we find also this behavior, cf.\
Fig.\ \ref{fig06}. In both cases we find that the HOMO is localized to both Si and Ge atoms,
whereas the LUMO is localized mainly to the Ge atoms. 

\section{conclusion}

In this paper, we have presented the results of our study of the structural and electronic
properties of naked (Si)Ge and (Ge)Si core/shell nanoparticles together with those of pure Si and Ge,
homogeneous SiGe and Si$\vert$Ge systems with a diameter of up to around 2 nm.
The interesting properties critically depend on the size of core and shell and also
on the type of core and shell atoms. Although we did not enter a detailed discussion of
this issue, we emphasize that our results on pure Si and Ge clusters, in particular concerning the 
reduced band gap compared with the crystalline material, are in excellent agreement with results of 
other experimental and theoretical studies. 

Due to difference in surface energy, (Si)Ge core/shell systems are more stable than are (Ge)Si
systems. Moreover, the stability of the (Si)Ge systems was quite well described in terms of 
the quantity $\Delta E_1$ as a function of the number of core shells, whereas $\Delta E_2$ 
as a function of the number of core shells gave a good description of the stability of 
the (Ge)Si systems. 

We did observe a marginal tendency of a Ge$\to$Si electron transfer, irrespectively of which
system was forming the core and which the shell. A similar effect was not found for the homogeneous SiGe
systems, but for the interface Si$\vert$Ge systems. 

In all systems, the (unpassivated) surface was dictating the properties of the frontier orbitals. 
The latter were localized to the surface, and their energies occurred in the energy gap of the 
pure, infinite crystals, leading to quite low band gaps. Therefore, surface passivation could be
a useful means of tuning the optical properties of these systems. In contrast to our earlier
results on naked AB semiconductor clusters, we do for the present ones not observe a correlation
between stability and band gap. Finally, for none of the systems we found a charge separation 
upon electronic excitation.

\begin{acknowledgments}
This work was supported by the German Research Council through project Sp439/11.
\end{acknowledgments}

\unitlength1cm

\unitlength1cm

\unitlength1cm

\unitlength1cm

\end{document}